\documentstyle[twoside,fleqn,espcrc2,psfig]{article}
\title{Effects of Chemical Potential on Hadron Masses at Finite Temperature
\thanks{presented by T.Takaishi}}
\author{ {\large QCD-TARO Collaboration:}
 Ph.~de~Forcrand${}^a$,      M.~Garc{\'\i}a~P\'erez${}^b$,
 T.~Hashimoto${}^c$,      S.~Hioki${}^d$,
 H.~Matsufuru${}^{e,f}$,    
 O.~Miyamura${}^e$,
 A.~Nakamura${}^g$,          I.-O.~Stamatescu${}^{f,h}$,
 T. Takaishi${}^i$
 and  T.~Umeda${}^e$ \\ 
\vspace{4mm}
${}^a$ SCSC, ETH-Z\"urich, CH-8092 Z\"urich, Switzerland  \\
${}^b$ Dept. F\'{\i}sica Te\'orica, Universidad Aut\'onoma de Madrid,
 E-28049 Madrid, Spain \\ 
${}^c$ Dept. of Appl. Phys., Fac. of Engineering,
        Fukui Univ., Fukui 910-8507, Japan  \\
${}^d$ Dept. of Physics, Tezukayama Univ.,
        Nara 631-8501, Japan  \\
${}^e$ Dept. of Physics, Hiroshima Univ.,
        Higashi-Hiroshima 739-8526, Japan  \\
${}^f$ Inst. Theor. Physik, Univ. of Heidelberg,
        D-69120 Heidelberg, Germany  \\
${}^g$ Res. Inst. for Inform. Sci. and Education, Hiroshima Univ.,
        Higashi-Hiroshima  739-8521, Japan  \\
${}^h$ FEST, Schmeilweg 5, D-69118 Heidelberg, Germany \\
${}^i$ Hiroshima University of Economics,
        Hiroshima 731-0192, Japan  \\ }

\begin{document}
\begin{abstract}
We study the effects of the chemical potential on the $\rho$ meson mass at
finite temperature. 
Our preliminary results show that some effects are seen in the vicinity 
of the phase transition point.   
Although the signal is still too noisy to obtain conclusive physical results 
within limited statistics, the mass susceptibility is consistent with zero.

\end{abstract}

\maketitle

\section{Introduction and Motivation}

QCD Sum Rule analysis has 
suggested that a mass-shift occurs in proportion 
to the baryon density of medium.
In nuclear matter, a few to 10 \% reduction of vector meson masses 
is expected \cite{HATSU} 
and confirmation of this fact  is of current experimental interest.  

Although the importance of a quantitative investigation on 
this subject is obvious, 
it seems difficult to tackle it by present lattice technology.\cite{BAR}  
A new attempt to measure the derivative of mass $M$
in terms of the chemical potential, $\partial M / \partial \mu$ at 
$\mu = 0$ has been made in this work.  
At $T=0$, since the density depends cubically on  $\mu$, 
$\partial M / \partial \mu$ vanishes. On the other hand , at finite 
temperature, a linear dependence on the chemical potential 
is expected and we will have a chance to see the density effect. 

\section{Formulation}
We use  2 flavors of the Wilson fermion for the fermionic part of the action.
The action is given by
$
S=S_G+S_F
$
where $S_G$ and $S_F$ stand for the gauge and the fermionic parts
respectively,
and $S_F$ has the form, 
\begin{equation}
S_F = \sum_{i=u,d} \bar{\psi}^i (n) D(n,m) \psi^i(m).
\end{equation}
For $S_G$ we use the standard Wilson action.
The Wilson matrix $D(n,m)$ with  chemical potential is written as\cite{HK,NAKA}
\begin{eqnarray}
D(n,m)& & =  \delta_{n,m} \\ \nonumber
& &  -  \kappa\sum_{i=1}^3 [(1-\gamma_i)U_i(m)\delta_{m,n+\hat{i}} \\ \nonumber
& & + (1+\gamma_i)U_i^\dagger(n-\hat{i})\delta_{m,n-\hat{i}}] \\ \nonumber
& & -\kappa [(1-\gamma_4)U_4(m)e^{\mu}\delta_{m,n+\hat{4}} \\ \nonumber
& & +(1+\gamma_4)U_i^\dagger(n-\hat{4})e^{-\mu}\delta_{m,n-\hat{4}}]
\end{eqnarray}

The hadron correlation function $C(t)$ is given by
\begin{eqnarray}
C(t) & = &
\label{eq1}
\int dU d\psi d\bar{\psi} H(t) H^{\dagger}(0)e^{-S} /Z \\
& = & \sum_i A_i \cosh(M_i(t-N_t/2)).
\label{eq2}
\end{eqnarray}
where $A_i$ may be written as $A_i=\tilde{A}_i/(2\sinh(M_i N_t/2))$.

To derive the expression of  the mass susceptibility on the chemical potential,
we take a derivative of $C(t)$ with respect to the
chemical potential $\mu$.
>From eq.(\ref{eq1}) one obtains
\begin{eqnarray}
\label{eq3}
\frac{\partial C(t)}{\partial \mu}=
& &-<H(t)H^{\dagger}(0)\frac{\partial S_F}{\partial \mu}>  \\ \nonumber
& &+ <H(t)H^{\dagger}(0)><\frac{\partial S_F}{\partial \mu}>.
\end{eqnarray}
On the other hand from eq.(\ref{eq2}) one obtains
\begin{eqnarray}
\label{eq4}
& &\sum_i [\frac{\partial A_i}{\partial \mu} \cosh(M_i(t-Nt/2) \\ \nonumber
& &-\frac{\partial m_i}{\partial \mu}A_i(t-Nt/2)
\sinh(M_i(t-Nt/2))].
\end{eqnarray}
Eqs.(\ref{eq3}) and (\ref{eq4}) should be equal to each other.
Eq.(\ref{eq3}) is evaluated by Monte Carlo simulations.
Eq.(\ref{eq4}) contains 4 unknown fitting parameters for a given
$i$.
$A_i$ and $M_i$ are determined from a fit to the hadronic
correlation
function of eq.(\ref{eq2}).
Then these results are used for eq.(\ref{eq4}) as fixed parameters and,
thus  the chemical potential dependent terms $\partial A_i/ \partial \mu$
and $\partial M_i/ \partial \mu$ are extracted from a two parameter
fit to eq.(\ref{eq4}).

\section{Calculation of $<H(t)H^{\dagger}(0)\frac{\partial S_F}{\partial
\mu}>$ }
In the present study we aim at obtaining the mass susceptibility at $\mu=0$.
At $\mu=0$, $<\frac{\partial S_F}{\partial \mu}>=0$.
Thus the calculation of eq.(\ref{eq3}) reduces to calculate
the remaining term $<H(t)H^{\dagger}(0)\frac{\partial S_F}{\partial \mu}>$.

Now we  specify the hadronic operator.
For the present study we take the $\rho$ meson operator,
which is given by,
\begin{equation}
H(t)=\sum_x u(x,t)\gamma_{\nu} \bar{d}(x,t).
\end{equation}
We take $\nu=2$ here.
The derivative of $S_F$ with respect to $\mu$ is
\begin{eqnarray}
&& \frac{\partial S_F}{\partial \mu}  \\ \nonumber
&& =  - \kappa \sum_{ i=u,d }
\sum_{n,m} \bar{\psi}^i (n)[(1-\gamma_4)U_4(m)e^{\mu}\delta_{m,n+\hat{4}}  \\
\nonumber
&&  -(1+\gamma_4)U_i^\dagger(n-\hat{4})e^{-\mu}\delta_{m,n-\hat{4}}]\psi^i(m) \\ \nonumber
&& \equiv   - \kappa \sum_{i=u,d}
\sum_{n,m} \bar{\psi}^i (n)[B_1 - B_2] \psi^i(m) , 
\end{eqnarray}
where
\begin{equation}
B_1=(1-\gamma_4)U_4 (m)e^{\mu}\delta_{m,n+\hat{4}} ,
\end{equation}
\begin{equation}
B_2=(1+\gamma_4)U_4^\dagger (n-4)e^{-\mu}\delta_{m,n-\hat{4}}.
\end{equation}

Let us consider
\begin{equation}
<H(t)H^\dagger(0)\kappa \sum_{n,m}\bar{u}(n)B_1u(m)>
\label{eq5}
\end{equation}
where $u(m)= \psi^u(m)$.
This term contains a connected and disconnected diagrams.
We neglect the disconnected part, which is unimportant for the present study.
Eq.(\ref{eq5}) is written as
\begin{eqnarray}
-\sum_{x,z}\gamma_\nu D^{-1}(0,0:z) (1-\gamma_4)U_4(z)\times && \\ \nonumber
 D^{-1}(z+\hat{4}:x,t) \gamma_\nu  D^{-1}(x,t:0,0). && 
\end{eqnarray}
where the chemical potential $\mu$ is set to zero.

Similar calculations are applied for the remaining terms ( for $B_2$ and
for $i=d$ ).

\begin{figure}[h]
\centerline{\psfig{figure=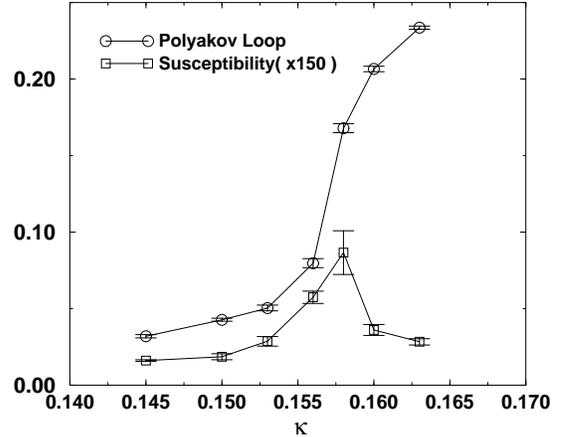,height=6cm}}
\vspace{-5mm}
\caption{Polyakov loop and its susceptibility as a function of $\kappa$.}
\end{figure}

\section{Simulation}
Simulations are done on a lattice size of $16\times8\times8\times4$ 
at $\beta=5.30$.

To find a finite temperature transition point,
the Polyakov loop and its susceptibility are
calculated, as shown in Fig.1
The finite temperature transition happens
around $\kappa=0.158$.
The chemical potential effect on hadron masses at $\mu=0$
is expected to start in the vicinity of the transition.
We take $\kappa=0.153$ and $\kappa=0.158$ for
the present study.
We use 30 configurations generated by the hybrid Monte
Carlo algorithm for each $\kappa$.
The configurations are separated by 40-50 trajectories.  

\section{Preliminary results}
Figs.2 and 3 show the derivative of the hadronic correlation function $C(x)$
with respect to  $\mu$ as a function of the largest lattice direction: $x$. The hadronic correlation function is 
measured in the x-direction.

\begin{figure}[ht]
\centerline{\psfig{figure=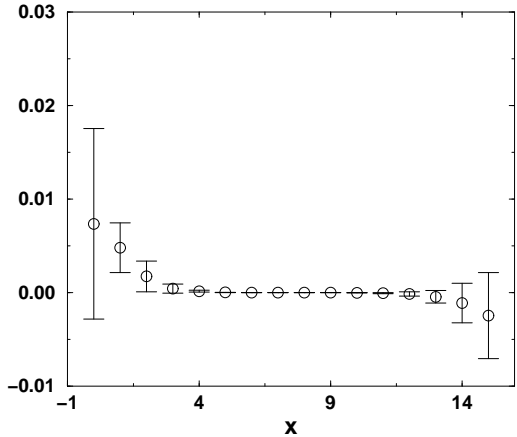,height=6cm}}
\vspace{-5mm}
\caption{The derivative of the hadronic correlator $C(x)$ with respect to
$\mu$ at $\kappa=0.153$ as a function of $x$.}
\end{figure}

\begin{figure}
\centerline{\psfig{figure=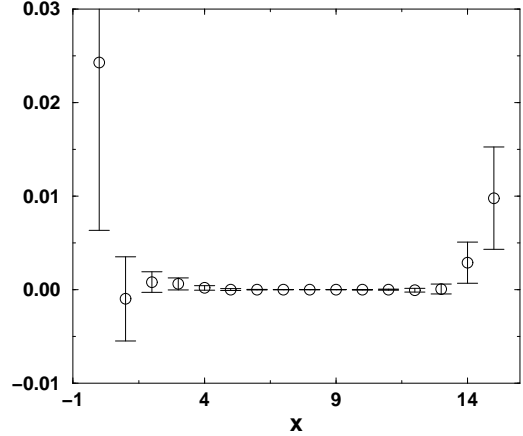,height=6cm}}
\vspace{-5mm}
\caption{Same as in Fig.2 but at $\kappa=0.158$.}
\end{figure}

Although the signals within the current statistics are still too noisy, we fit
the data using eqs.(\ref{eq3}) and (\ref{eq4}).  The mass susceptibility,
$\partial M/\partial \mu$, is consistent with zero, while 
$\partial \tilde{A} / \partial \mu$ could be finite and positive.

\section{Discussions}

The effect of chemical potential on the mass susceptibility 
is suppressed by $\exp(-M_B/T)$ where $M_B$ is the baryon mass. 
Our results may show that $M_B$ at our simulation parameter is too heavy
to obtain a reasonable signal within limited statistics.
Our hope is that signal may be enhanced if a small quark mass is used.

One of the authors (T.T.) would like to thank
Electric Technology Research Foundation of Chugoku for financial support.
Calculations reported here were done on AP1000 at Fujitsu Parallel 
Computing Research Facilities and SX-4 at RCNP, Osaka Univ.


\begin{thebibliography}{9}
\bibitem{BAR} I.M.Barbour, S.E.Morrison, E.G.Kepfish, J.G.Kogut and M-P.Lombardo,
Nucl. Phys. B (Proc. Suppl.) 60A (1998) 220
\bibitem{HATSU}T.Hatsuda and S-H.Lee, Phys. Rev. C46 (1992) 34.
\bibitem{HK} P.Hasenfratz and  F.Karsch, Phys. Lett. 125B (1983) 308.
\bibitem{NAKA} A.Nakamura, Phys. Lett. 149B (1984) 391, Acta Physica Polonica B16 (1985) 635
\end{thebibliography}
\end{document}